\documentclass[superscriptaddress,groupedaddress,nofootnoteinbib,11pt]{article}

\usepackage{jcappub}
\usepackage{bm}
\usepackage{verbatim}

\usepackage{epsf}
\usepackage{graphicx,epsfig}
\usepackage{amsfonts}
\usepackage{amssymb}
\usepackage{xcolor}

\definecolor{mine}{rgb}{0.2,0.1,0.7}
\definecolor{bb}{rgb}{0.3, 0.5, 1}
\definecolor{bg}{rgb}{0.1, 0.1, 0.5}

\def\L{\Lambda}

\def\d{\mathrm{d}}

\def\L*{{\cal L}_*}
\def\L{\mathcal{L}}
\def\({\left(}
\def\){\right)}

\def\nn{\nonumber}

\def\p{\partial}

\def\<{\langle}
\def\>{\rangle}

\newcommand{\bea}{\begin{eqnarray}}
\newcommand{\eea}{\end{eqnarray}}
\newcommand\be{\begin{equation}}
\newcommand\ee{\end{equation}}
\newcommand\beq{\begin{equation}}
\newcommand\eeq{\end{equation}}

\def\ba{\begin{eqnarray}}
\def\ea{\end{eqnarray}}

\def\m{\mu}
\def\n{\nu}
\def\p{\phi}

\def\Cone{T_1}
\def\Ctwo{T_2}
\def\Cthree{T_3}
\def\Cfour{T_4}
\def\Cfive{T_5}
\def\Csix{T_6}
\def\Ceight{T_7}
\def\Cnine{T_8}
\def\Cten{T_9}

\def\Vone{{\cal O}_1}
\def\Vtwo{{\cal O}_2}

\def\Ffour{F_4}
\def\FfourX{F_{4,X}}
\def\FfourXX{F_{4,XX}}
\def\FfourXXX{F_{4,XXX}}

\def\Ffive{F_5}
\def\FfiveX{F_{5,X}}
\def\FfiveXX{F_{5,XX}}
\def\FfiveXXX{F_{5,XXX}}

\def\c{c_s}

\newcommand{\refeq}[1]{(\ref{#1})}

\begin{document}

\title{Non-Gaussian inflationary shapes \\
in $G^3$ theories beyond Horndeski}

\author[1]{Matteo Fasiello}
\author[2,3]{and S\'ebastien Renaux-Petel} 
  \affiliation[1]{CERCA \& Department of Physics, Case Western Reserve University, 10900 Euclid Ave, Cleveland, OH 44106, USA}
  \affiliation[2]{Laboratoire de Physique Th\'eorique et Hautes
   Energies, Universit\'e  Pierre \& Marie Curie - Paris VI, CNRS-UMR 7589, 4 place Jussieu, 75252 Paris, France}
\affiliation[3]{Sorbonne Universit\'es, Institut Lagrange de Paris,
  98 bis Bd Arago, 75014 Paris, France}
\emailAdd{matte@case.edu}
\emailAdd{srenaux@lpthe.jussieu.fr}


\vskip 8pt

\date{\today}


\abstract{ We consider the possible signatures of a recently introduced class of healthy theories beyond Horndeski models on higher-order correlators of the inflationary curvature fluctuation. Despite the apparent large number and complexity of the cubic interactions, we show that the leading-order bispectrum generated by the \textit{Generalized Horndeski} (also called $G^3$) interactions can be reduced to a linear combination of two well known $k$-inflationary shapes. We conjecture that said behavior is not an accident of the cubic order but a consequence dictated by the requirements on the absence of Ostrogradski instability, the general covariance and the linear dispersion relation in these theories.}

\maketitle

\section{Introduction}

Whenever in search of phenomenologically viable theories of the early universe, one in confronted with the possibility  to walk a number of intrinsically different paths. Simplicity might well be the guiding principle; on the other hand, a complementary approach suggests that there's much to be learned and gained by working in full generality. An effective field theory (EFT) approach \cite{Creminelli:2006xe,Cheung} belongs to the latter perspective and has recently received considerable attention, which has lead to a widespread effort in the current literature to employ it in just about every realm of cosmology (see \textit{e.g.} \cite{Baumann:2010tm,Carrasco:2012cv,Bartolo:2010bj,Gubitosi:2012hu,Hertzberg:2012qn,Pajer:2013jj,Mercolli:2013bsa,Piazza:2013coa,Porto:2013qua,Baldauf:2014qfa,Tsujikawa:2014mba}). 

What we will be concerned with in this paper is yet another perspective which stands in between the simplicity-\textit{vs}-generality dichotomy. The starting point will be the EFT of Ref.~\cite{Gleyzes:2013ooa}, which is itself an application to dark energy theories of the ``effective field theory of fluctuations" paradigm first introduced in \cite{Creminelli:2006xe,Cheung}, augmented by the requirement that the Lagrangian in the so-called unitary gauge generates equations of motion (eom's) that are at most second order in (time) derivatives for linear perturbations. This last requirement is a sufficient (but not a necessary one) condition to guarantee the absence of so called Ostrogradski instabilities \cite{Ostro}, at least perturbatively. 

 In the pioneering work of Horndeski  \cite{Horndeski:1974wa} one can already find a large class of Lagrangians that \textit{automatically} generate safe second order eom's. However, as has been known for a while \cite{Blas:2009yd,deRham:2011qq,Chen:2012au,Zumalacarregui:2013pma}, the apparent order of the eom's does not always reflect the number of  healthy effective degrees of freedom. Indeed, in the presence of (sometimes hidden) constraints one may show that the \textit{propagating} degrees of freedom, the ones that matter, do indeed have, after some work, second order equations of motion. This is certainly the case of the generalized Horndeski, also called $G^3$ theories  \cite{Gleyzes:2014dya,Gleyzes:2014qga}, where the tracking down of gauge redundancies has made it possible to go beyond the Horndeski Lagrangian adding to it new and well-behaved terms. Intriguingly, these new pieces generate an interesting phenomenology. It is the case for example of the speed of propagation of matter that, contrary to the standard picture, is now affected by the presence of the scalar degree of freedom even when matter is only minimally coupled to gravity (see \textit{e.g.} Refs.~\cite{Gleyzes:2014dya,Gleyzes:2014qga,Kase:2014yya}). 
 
What motivates our analysis here however are the signatures that the generalized Horndeski terms might reveal if employed as leading interactions in a generalized Hornedski Lagrangian for the inflaton field (a similar motivation prompted several interesting analyses in the case of the, by now standard, Horndeski theory \cite{Kobayashi:2010cm,Kobayashi:2011nu,Gao:2011qe,DeFelice:2011uc,RenauxPetel:2011sb,DeFelice:2013ar}). The quadratic action for cosmological fluctuations in the $G^3$ theory has been calculated in Ref.~\cite{Gleyzes:2014dya} and turn out to be perfectly mundane. Hence, no specific signature is expected at the level of the primordial power spectrum. The characterization of higher-order correlation functions is therefore of paramount importance if we are to observationally probe the inflationary dynamics of these scenarios. We concentrate in this paper on the primordial bispectrum and perform our analysis in the regime where one can safely neglect metric perturbations. The latter are suppressed by the parameter $\epsilon_{mix}\sim E_{mix}/H$, where $E_{mix}$ is an energy scale determined from the normalization of the leading kinetic term in the quadratic action for the $\delta \phi$-generated part of the observable $\zeta$. In the language of Refs.~\cite{Creminelli:2006xe,Cheung}, this is the scale at which the dynamics of the Goldtsone $\pi$ decouples from the metric and the typical value for $\epsilon_{mix}$ is $ |\dot{H} |^{1/2}/H$, where $H$ is the Hubble parameter.

 Intriguingly, we find that the new interactions of $G^3$ theories generate, out of the numerous interaction terms, a non-Gaussian signal that is extremely simple and is comprised by the shapes of solely two well-known operators of the curvature perturbation $\zeta$, namely $\dot{\zeta}^3$ and $\dot \zeta (\partial_i \zeta)^2$. This drastic simplification has been obtained by the judicious uses of integration by parts and of the linear equation of motion for $\zeta$. The latter exact procedure enables one to eliminate redundancies in the basis of interaction operators generating the non-Gaussian signal and have already been put to good use in the class of Horndeski theories, where similar simplifications arise \cite{RenauxPetel:2011sb}. Our result, which does not hold for the most generic effective field theory of Ref.~\cite{Creminelli:2006xe,Cheung}, might well be tied to the highly specific structure of the extended Horndeski Lagrangian and therefore be related to the absence of Ostrogradski instabilities.

The paper is organized as follows: in \textit{Section 2}, after a brief discussion of Ostrogradski instabilities,
we provide an overview of the extended Horndeski theories of Ref.~\cite{Gleyzes:2014dya} and employ them as the Lagrangian describing the inflaton field. 
In \textit{Section 3} we study in detail the non-Gaussian imprints of the new interactions beyond Horndeski and show how solely two operators encode the primordial bispectrum they generate. In \textit{Section 4}  we offer comments on our results, discuss the context in which they emerged and point interesting venues one could pursue. An appendix offers a consistency check of our calculations.

 \section{Inflating with Generalized Horndeski}

\subsection{A brief clarification on Ostrogradski instabilities}

It is generally best to analyze the degrees of freedom and the health of a given theory when the latter has been put in Hamiltonian form (see \textit{e.g.} Ref.~\cite{Dirac}). It is in this context that Ostrogradski has shown how, whenever a (non degenerate) theory possesses equations of motion that are beyond second order and those eom's are not complemented by enough constraints that appropriately reduce the dimensionality of phase space, instabilities inevitably arise \cite{Ostro}. A theory with higher order eom's necessarily requires more than two initial conditions. This corresponds to a higher number of modes which participate in the dynamics and a larger phase space. Some of the canonical variables appear linearly in the Hamiltonian, thus making it possible to eventually generate and excite modes of arbitrarily negative energy\footnote{Note that, for this to happen, the Hamiltonian need not be unbounded from below \cite{Woodard}.}, and making the system unstable.

Note that these instabilities are not always, necessarily, a problem. Classically for example, as far as a theory is not an interacting one, such a system will not excite arbitrarily negative energy modes\footnote{The interest in such theories is, of course, quite limited.}. Much more importantly, a full effective field theory approach can, under specific circumstances, deal with higher order derivative interactions in a rather straightforward fashion: if the derivative expansion is organized around a well-defined perturbative expansion parameter, one can employ the second order equations of motion to handle, order by order, the higher derivative interaction terms. An example of this familiar procedure is given in the work \cite{Weinberg:2008hq} and was analyzed in greater detail in Ref.~\cite{JLM}.  The validity of the procedure goes at least as far as the scale at which the effective theory description is itself valid. 

Let us stress here that this process is different from our use of the second-order equation of motion in \textit{Section} \ref{sec:simplification} of this paper: in our case, there is absolutely no Ostrogradski instability, even at very high energies. One could well perform the calculation of the primordial bispectrum using the complicated form Eqs.~\refeq{L4cubic}-\refeq{L5cubic} of the cubic action, without simplifying it with the linear equation of motion. However, by proceeding in this way, one could miss redundancies between operators and erroneously infer the existence of new non-Gaussian shapes, as one can easily encounter in the literature.

\subsection{$G^3$ theories as the inflaton Lagrangian}
An almost flat potential is all that is necessary for a successful inflationary mechanism\footnote{An almost flat potential is sufficient to produce a long enough period of inflation and to generate an almost scale-invariant primordial power-spectrum. However, one should always bear in mind that the couplings of the inflaton to the Standard Model degrees of freedom should ultimately be prescribed in order to successfully connect the inflationary era to the radiation era through the period of (p)reheating.}. Such a scenario is of course included, among many others, if the Horndeski Lagrangian is elected to be the inflationary one. On the other hand, Hornedski theories span a vastly richer phenomenology offering distinct signatures (see \textit{e.g.} Refs.~\cite{Barrow:2012ay,Koyama:2013paa,Gomes:2013ema,Kase:2013uja,Bettoni:2013diz,Deffayet:2013lga,Jimenez:2013qsa,Nishi:2014bsa,Charmousis:2014zaa,Charmousis:2014mia}) and, this class having just been enlarged by \textit{Generalized Horndeski} terms \cite{Gleyzes:2014dya}, one ought to explore possible further novelties and strive for a complete characterization. 

Obtaining a successful inflating background in the case at hand is a simple task because the new terms do not change the qualitative feature of the FRW solution. The analysis has been performed in a number of works in the recent literature. Instead of reproducing a remarkably similar content here, we offer some general comments which hold true for all increasedly-generalized Horndeski models. Following Ref.~\cite{Gleyzes:2014dya}, we write the generalized Horndeski Lagrangian (henceforth also the inflationary Lagrangian) as $L=\sum_{a=2}^{5}L_a $
with
\bea
L_2&=& A_2  \label{A2}  \\
L_3 &=& (C_3 + 2 X C_{3,X}) \Box \phi + X C_{3, \phi}  \\
L_4 &=& B_4\, R - \frac{A_4 +B_4}{X} \big[(\Box \phi)^2 - \phi_{\mu \nu} \phi^{\mu \nu}\big] +(C_4 + 2 X C_{4,X}) \Box \phi + X C_{4, \phi} \nonumber \\
 &+&  2 \frac{A_4+B_4 - 2 X B_{4,X}}{X^2}(\phi^{\mu} \phi^{\nu} \phi_{\mu \nu} \Box \phi - \phi^{\mu}  \phi_{\mu \nu} \phi_{\lambda} \phi^{\lambda \nu})    \label{L44}  \\
L_5&=&G_5 \,G_{\mu \nu}\phi^{ \mu \nu} 
- (-X)^{-3/2} A_5 \big[ (\Box \phi)^3 - 3 (\Box \phi) \phi_{\mu \nu} \phi^{\mu \nu}+ 2 \phi_{\mu \nu} \phi^{\nu \rho} \phi^\mu_{\ \rho}\big]  \nonumber \\
 &-& \frac{X B_{5,X} + 3  A_5}{(-X)^{5/2}}\big[ (\Box \phi)^2 \phi_\mu \phi^{\mu \nu} \phi_\nu - 2 \Box \phi \phi_\mu \phi^{\mu \nu} \phi_{\nu \rho} \phi^\rho 
- \phi_{\mu \nu} \phi^{\mu \nu} \phi_\rho \phi^{\rho \lambda} \phi_\lambda+ 2 \phi_\mu \phi^{\mu \nu} \phi_{\nu \rho} \phi^{\rho \lambda} \phi_\lambda \big] \nonumber \\
 & +& C_5 \, R - 2 C_{5,X} \, \big[ (\Box \phi)^2 - \phi^{ \mu \nu} \phi_{ \mu \nu} \big] +(D_5 + 2 X D_{5,X}) \Box \phi + X D_{5, \phi} \,\, ,\,\,\,  \,\,  \label{L55}
\eea
where $\phi_{\mu}\equiv \nabla_{\mu} \phi$, $\phi_{\mu \nu} \equiv \nabla_{\mu} \nabla_{\nu} \phi$, $X \equiv g^{\mu \nu} \phi_{\mu} \phi_{\nu}$, the quantities $A_n, B_n(\phi,X)$ are generic functions, $R$ and $G_{\mu \nu}$ denote respectively the Ricci scalar and the Einstein tensor of the metric tensor $g_{\mu \nu}$, and:
\bea
C_3&\equiv& \frac{1}{2}\int  A_3  (-X)^{-3/2}   dX  \,; \quad  C_4 \equiv - \int  B_{4,\phi} (-X)^{-1/2} dX\, ; \quad  G_{5} \equiv - \int  B_{5,X}(-X)^{-1/2} dX\, ;\,\,  \nonumber \\
  \quad  C_5 &\equiv&  -\frac14 X \int  B_{5,\p}(-X)^{-3/2} dX\, ; \quad D_5 \equiv  - \int  C_{5,\p}(-X)^{-1/2} dX\,  \nn .
\eea
The above Lagrangian reduces to (combinations of) the Horndeski one only if $A_{4}$ and $A_5$ are given in terms of $B_4$ and $B_5$ by
\be
A_4=-B_4+2 X B_{4,X}\,, \qquad A_5=-X B_{5,X}/3\,,
\ee
under which conditions the second lines of Eqs.~\refeq{L44} and \refeq{L55} vanish. This implies that the generalized theory under scrutiny contains two additional free functions besides Horndeski's ones, and that we can rewrite the total action in the form:
\be
S = \int \d^4 x \, \sqrt{-g} \left({\cal L}_{{\rm Horndeski}}+ \L_4+\L_5  \right)
\ee
where
\be
\L_4=\Ffour(\phi,X) \left( \p^\m \p^\n \p_{\m \n} \Box \p -\p^\m \p_{\m \n} \p^{\n \lambda}  \p_\lambda -\frac{X}{2} \left( (\Box \p)^2-\p_{\m \n} \p^{\m \n} \right) \right)\,,
\label{L4}
\ee
\bea
\L_5&=&\Ffive(\phi,X) \left((\Box \p)^2 \p^\m \p^\n \p_{\m \n}  -2 \Box \p\, \p^\m \p_{\m \n} \p^{\n \lambda}  \p_\lambda -(\p_{\m \n} \p^{\m \n})(\p^\lambda \p^\rho \p_{\lambda \rho} ) +2 \p^\m \p_{\m \n} \p^{\n \rho} \p_{\rho \lambda} \p^{\lambda}
 \right.
\nn
\\
&& 
\left. 
-\frac{X}{3} \left( (\Box \p)^3 -3 \Box \p \p_{\m \n} \p^{\m \n}+2 \p_{\m \n} \p^{\n \rho} \p^{\m}_{\,\,\rho}  \right)
  \right)\,,
\label{L5}
\eea 
and $\Ffour$ and $\Ffive$ are generic free functions of the inflaton field $\phi$ and its kinetic term $X$ (note that when $F_4$ and $F_5$ are constants, $\L_4$ and $\L_5$ boil down to the simple covariantizations of the original Galileon Lagrangians \cite{Nicolis:2008in}). Equivalently, one can perform integrations by part to find the expressions
\be
\L_4 = \left( 2 \Ffour +X \FfourX  \right) \left( \p^\m \p^\n \p_{\m \n} \Box \p -\p^\m \p_{\m \n} \p^{\n \lambda}  \p_\lambda \right)  -\frac{X}{2} \Ffour R_{\mu \nu}\p^{\m} \p^{\n} -\frac{X}{2}F_{4,\phi} \left(\p^\m \p^\n \p_{\m \n}-X \Box \p  \right)  \,,
\label{L4bis}
\ee
\bea
\L_5&=& \frac13 \left(5 \Ffive +2 X \FfiveX  \right) \left((\Box \p)^2 \p^\m \p^\n \p_{\m \n}  -2 \Box \p\, \p^\m \p_{\m \n} \p^{\n \lambda}  \p_\lambda -(\p_{\m \n} \p^{\m \n})(\p^\lambda \p^\rho \p_{\lambda \rho} )   \right.
\nn
\\
&& 
\left. +2 \p^\m \p_{\m \n} \p^{\n \rho} \p_{\rho \lambda} \p^{\lambda}  \right)+ \frac13 X F_{5,\phi} \left( X \left( (\Box \p)^2-\p_{\m \n} \p^{\m \n} \right)-2 \Box \p  \p^\m \p^\n \p_{\m \n}+2 \p^\m \p_{\m \n} \p^{\n \lambda}  \p_\lambda \right) \nn \\
&+&\frac23 X \Ffive \left(R_{\sigma \mu \rho \nu} \p^{\mu} \p^{\rho \sigma} \p^{\n}+R_{\m \n} \p_{\sigma} \p^{\m \sigma} \p^{\n}-R_{\m \n} \p^{\m} \p^{\n} \Box \p \right) \,.
\label{L5bis}
\eea
Despite the fact that the Lagrangians \refeq{L4} and \refeq{L5} do not belong to the class of Horndeski's ones, and therefore generate higher-order equations of motion, they do propagate only three degrees of freedom (one scalar mode, plus the two standard tensor modes). Not obvious in this language, this becomes transparent by resorting to the uniform inflaton gauge, in which the Lagrangian depends only on the metric and its \textit{first} derivatives (see Refs.~\cite{Gleyzes:2014dya,Lin:2014jga,Gleyzes:2014qga} for Hamiltonian analyses). Furthermore, the absences of ghosts and gradient instabilities are guaranteed as long as two mild conditions on the $A_n,B_n$ are satisfied.\\

The (generalized) Horndeski Lagrangian contains higher derivative terms. From this alone we know from dimensional analysis that a scale is introduced and that above a certain energy the higher derivative Horndeski interactions become important for the inflationary dynamics. This is most striking in the (covariant) Galileon limit of the generalized Lagrangian \cite{Burrage:2010cu}\footnote{In flat space one recovers Galileon terms by setting $B_4=0=B_5$ ,  $A_4=-X^2\,, \, A_5=(-X)^{5/2}\, $. These theories, ubiquitous in the literature, have been first derived in a much different context \cite{Dvali:2000hr}.} as the number of derivatives per scalar ends up being always equal to or higher than the corresponding number in the more general case. This means that there is an energy regime in which the dynamics is really probing the presence of the Horndeski interactions and it is in such a regime one ought to search for their imprints.

As well known, Galileon theories are, at least in flat space, invariant under the Galileon symmetry and, consequently, under shift. These properties make them particularly compelling in that, as a result, the coefficients of the Galileon interactions are not renormalized. One must eventually break the Galileon symmetry in order to realize inflation but this can be done in a controlled way which allows the Galileon inflationary theory to inherit approximate non-renormalization.

The more general Horndeski theories do not share this last property but are nevertheless stable at the fully non-linear level --- that is, they do not involve Ostrogradski instabilities --- and, as mentioned, the latest additions to the family (\textit{i.e.} the Lagrangians Eqs.~\refeq{L4}-\refeq{L5} above) have particularly interesting consequences on the speed of propagation of matter. Given these appealing properties and the motivations mentioned above, we study their inflationary signatures in the next section.

\section{One Number Says It All}
\label{sec:simplification}

\subsection{Set-up and third-order action}

The aim of this section is to characterize the non-Gaussian inflationary signal generated by the generalized Horndeski interactions in Eqs.~\refeq{L4} and \refeq{L5}. In the following, we shall assume that the background evolution of this system is such that an almost de-Sitter inflationary phase is achieved, and we study the behavior of the scalar perturbation about this background. The linear analysis performed in Ref.~\cite{Gleyzes:2014dya} reveals that the scalar second-order action takes the form
\be
S_{(2)}= \int \d t \, \d^3 x\, a^3 \alpha \left( \dot \zeta^2-\c^2 \frac{(\partial_i \zeta)^2}{a^2}  \right)\,,
\label{S2}
\ee
where the explicit expressions of $\alpha$ and of $\c^2$, which are not important for our analysis, can be found in \cite{Gleyzes:2014dya}. Here, $\zeta$ is the gauge-invariant scalar curvature perturbation, which, at linear order, reads
\be
\zeta=\psi+\frac{H}{\dot {\bar \phi}} Q\,,
\ee
where the field $\phi$ is decomposed as $\phi=\bar \phi(t)+Q(t,x^i)$ and $\delta g_{ij}=-2 a^2 \psi\, \delta_{ij}$. 

Note that the form of the second-order action \refeq{S2} is completely standard: the absence of higher time derivatives in Eq.~\refeq{S2} is a direct manifestation of the fact that the theory possesses only one propagating degree of freedom, while the absence of higher spatial derivatives is a build-in requirement of the construction of the theory \cite{Gleyzes:2013ooa,Gleyzes:2014dya}.\\

In the following, we work at leading order in a generalized slow-roll approximation, that is we consider that all relevant quantities (like $H$, $\dot {\bar \phi}$, $\alpha$ and $\c$) evolve much less rapidly than the scale factor, $\dot X  /(HX) \ll 1$, so that they can be considered as constant for all practical purposes. Equivalently, we assume that the inflationary sector enjoys an approximate shift-symmetry\footnote{As well documented \cite{Kobayashi:2011nu}, one can certainly end inflation whilst an approximate shift symmetry is in place.}, so that derivatives of $\Ffour$ and $\Ffive$ with respect to $\phi$ can be neglected. Such a symmetry is a desirable feature in that a shift-symmetric theory has a de Sitter solution as an attractor \cite{Kobayashi:2011nu} and may be protected against large quantum corrections.
As we have explained in the introduction, we also concentrate on the leading-order non-Gaussianities, \textit{i.e.} we neglect the mixing with gravity. Under these assumptions, the linear equation of motion deduced from Eq.~\refeq{S2} reads
 \bea
\ddot Q+3H \dot Q -c_s^2 \partial^2 Q=0\,,
\label{linear-eom}
 \eea
 which we will use abundantly in the rest of this paper.\\
 
 To characterize the bispectrum signal generated by the new interactions Eqs.~\refeq{L4bis}-\refeq{L5bis}, the first step is to calculate the corresponding cubic action (as a consistency check, we perform the calculation starting from the form Eqs.~\refeq{L4}-\refeq{L5} in the appendix \ref{other-calculation}). Decomposing $\phi$ into its background plus fluctuating part, a long but straightforward calculation yields (here and in what follows, we omit the bar on the various background quantities unless an ambiguity can arise)\footnote{Note that the appearance of terms in $\ddot Q$ does not contradict the fact that there is only one propagation degree of freedom in this theory. Terms in Eqs.~\refeq{L4cubic}-\refeq{L5cubic} with second order time derivatives of $Q$ can always be put in a form, through integrations by part, that render them manifestly first order in time derivatives.}: 
 \bea
\frac{\L_{4 ({\rm cubic})} }{\Ffour \dot \phi}&=&   \left(2+\frac{X \FfourX}{\Ffour}  \right) \left[  -2( \Cone \Cfour- \Csix) -3 H  \Cthree+H  \Ctwo \Cfour \right]   \nonumber  \\
&+& 3H \left(2+4\frac{X \FfourX}{\Ffour} +\frac{X^2 \FfourXX}{\Ffour}\right) \left[ 4  \Cone \dot Q -H \Ctwo \dot Q  \right]  \nonumber \\
&+&  \left(2+10 \frac{X \FfourX}{\Ffour}  +3 \frac{X^2 \FfourXX}{\Ffour} \right) H \Ctwo \ddot Q + \left(4+8\frac{X \FfourX}{\Ffour} +2\frac{X^2 \FfourXX}{\Ffour}\right) \left[ \Cfour \dot Q \ddot Q- \Cfive \dot Q \right]   \nn \\
&-&  \left(6+24 \frac{X \FfourX}{\Ffour}  +15 \frac{X^2 \FfourXX}{\Ffour}+2  \frac{X^3 \FfourXXX}{\Ffour} \right)  \left[ 3H \dot Q^2 \ddot Q +H^2  \dot Q^3   \right]
\label{L4cubic}
\eea
and
\bea
\frac{3\, \L_{5 ({\rm cubic})} }{\Ffive \dot \phi^2}&=& 8 H   \left(5+2\frac{X \FfourX}{\Ffour}  \right)  ( \Cone \Cfour - \Csix) +2 \left(13+6\frac{X \FfourX}{\Ffour}  \right) H^2 \Cthree  -4 H^2   \left(3+\frac{X \FfourX}{\Ffour}  \right)      \Ctwo \Cfour  \nn \\
&-&  2  \left(12+ 25 \frac{X \FfiveX}{\Ffive} +6  \frac{ X^2 \FfiveXX}{\Ffive} \right) H^2 \Ctwo \ddot Q -4  \left(39+56 \frac{X \FfiveX}{\Ffive} +12  \frac{ X^2 \FfiveXX}{\Ffive} \right)  H^2 \Cone \dot Q \nn \\
&+&2  \left(3+2 \frac{X \FfiveX}{\Ffive}  \right)  H^3 \Ctwo \dot Q -4  \left(6+ 9 \frac{X \FfiveX}{\Ffive} +2  \frac{ X^2 \FfiveXX}{\Ffive} \right)    H^2 \Cfour \dot Q^2 \nn \\
&-& 4  \left(15+ 20 \frac{X \FfiveX}{\Ffive} +4  \frac{ X^2 \FfiveXX}{\Ffive} \right) H   \left(   \Cfour \dot Q \ddot Q - \Cfive \dot Q \right) 
 \nonumber \\
&+& 6 \left(30+75 \frac{X \FfiveX}{\Ffive}  +36 \frac{ X^2 \FfiveXX}{\Ffive}+4 \frac{ X^3 \FfiveXXX}{\Ffive} \right)  H^2 \left( \dot Q^2 \ddot Q+2 H \dot Q^3  \right)  \nonumber \\
&+&  \left(5+2  \frac{X \FfiveX}{\Ffive}  \right) \left[( \Cfour^2-\Ceight) \ddot Q - 2 \left( \Cfour \Cfive - \Cnine \right) \right] \,,\label{L5cubic}
\eea
where we use the following short-hand notations for spatially covariant combinations of derivatives of $Q$: 
\bea
a^2 \Cone &= &Q^{,i} \dot Q_{,i} \\
a^2 \Ctwo &=&  (\partial_i Q)^2  \\
a^4 \Cthree &=& Q^{,i}  Q^{,j} Q_{,ij} \\
a^2 \Cfour &=& \partial_i^2 Q  \\
a^2 \Cfive&=&   (\partial_i \dot Q)^2  \\
a^4 \Csix&=& \dot Q^{,i}  Q^{,j} Q_{,ij} \\
a^4 \Ceight&=& Q^{,ij} Q_{ij}\\
a^4 \Cnine &=& \dot Q^{,i} \dot Q^{,j} Q_{,ij}\,.
\eea

\subsection{Simplification of the third-order action}
\label{simplify}

At first sight, the high number of cubic interactions present in Eqs.~\refeq{L4cubic}-\refeq{L5cubic}, as well as their intricate expressions (involving for instance second spatial derivatives), can certainly make one think that new shapes of non-Gaussianities beyond Horndeski appear in this set up. This is indeed the case if each operator is considered independently from the others. However, the stability requirement of the Generalized Hornderski theory comes hand in hand with a specific structure at the cubic level: see the appearance of the combinations $\Cone \Cfour- \Csix$ and  $\Cfour \dot Q \ddot Q - \Cfive \dot Q$ in Eqs.~\refeq{L4cubic}-\refeq{L5cubic}, and the last line of Eq.~\refeq{L5cubic}. Therefore, before embarking oneself to calculate the three-point correlation function of each operator and draw hasty consequences, it is useful to pause and think of which simplifications might already occur at the level of the action. As we have announced already, this will lead to drastic and important simplifications. For this reason, we give below in some detail the various steps that allow us to simplify the cubic action. We define the two operators
\bea
\Vone \equiv \dot Q^3\, \quad {\rm and} \quad \Vtwo \equiv \dot Q \Ctwo=\dot Q (\partial_i Q)^2/a^2\,. \label{Standard-operators}
\eea

\noindent \underline{\bf{Integrations by part:}}\\

\noindent The simplest simplifications arise due to mere integrations by part. The relation 
 \bea
 \int \d t \, \d^3 x\, a^3 \dot Q^2 \ddot Q= -\int \d t \, \d^3 x\, a^3 H \Vone 
 \eea
 is straightforward. Simple spatial and temporal integrations by part also give
 \bea
 \int \d t \, \d^3 x\, a^3 \left(\Cone \Cfour- \Csix \right)= -\int \d t \, \d^3 x\, a^3 H \Cthree \,,
 \label{C1C4-C6}
 \eea
 while one simply has
 \bea
 \int \d t \, \d^3 x\, a^3 \Cthree=-\frac12\int \d t \, \d^3 x\, a^3 \Ctwo \Cfour \,. \label{C2C4}
\eea
\noindent \underline{\bf{Redundancy and use of the linear equation of motion:}}\\

\noindent Much less trivial simplifications arise due to the redundancy of some operators: as one of us highlighted in the context of Horndeski theories \cite{RenauxPetel:2011sb}, it is legitimate to use the linear equations of motion to simplify the interacting Lagrangian (see also Refs.~\cite{Seery:2005gb,Seery:2006tq,Seery:2010kh,Arroja:2011yj,Burrage:2011hd}). This simply stems from the fact the evaluation of higher-order correlators is made by using the propagators deduced from the second-order action, and that $\delta S_{(2)}/\delta Q=0$ by construction when evaluated on a propagator. This procedure, which is not accompanied by any field redefinition, is thus exact and valid at any perturbative order.\\

 \noindent  \textbullet \, By using the linear equation of motion Eq.~\refeq{linear-eom}, one finds two well known redundancies (see \textit{e.g.} Refs.~\cite{Creminelli:2010qf,Burrage:2010cu,RenauxPetel:2011dv,RenauxPetel:2011uk}):
 \bea
  \int \d t \, \d^3 x\, a^3 \Cfour \dot Q^2 &=&   \int \d t \, \d^3 x\, a^3 \dot Q^2 \frac{\partial^2 Q}{a^2}= 2  \int \d t \, \d^3 x\, a^3 \frac{H}{c_s^2} \Vone \label{simplification1}\,,  \\
\int \d t \, \d^3 x\, a^3 \Ctwo \Cfour  &=& 2 \int \d t \, \d^3 x\, a^3 \frac{H}{c_s^4} \left(\Vone+c_s^2 \Vtwo \right) \label{simplification2}\,.
 \eea
 From Eqs.~\refeq{C1C4-C6}-\refeq{C2C4} and \refeq{simplification2}, one can thus replace the three operators $\Cone \Cfour-\Csix$, $\Cthree$ and $\Ctwo \Cfour$ by a linear combination of the standard operators $\Vone$ and $\Vtwo$, and from Eq.~\refeq{simplification1}, one can replace $ \Cfour \dot Q^2$ by its expression in terms of $\Vone$. For the other operators:\\
 
  \noindent  \textbullet \, Replacing $\ddot Q$ in $\Ctwo \ddot{Q}$ by using the linear equation of motion Eq.~\refeq{linear-eom}, and using Eq.~\refeq{simplification2}, readily gives
 \bea
 \int \d t \, \d^3 x\, a^3 \Ctwo \ddot Q= \int \d t \, \d^3 x\, a^3 \frac{H}{c_s^2} \left(2 \Vone- c_s^2 \Vtwo \right)\,.
 \eea
   \noindent  \textbullet \, Spatially integrating by part, and using Eq.~\refeq{linear-eom}, one easily obtains
 \bea
 \int \d t \, \d^3 x\, a^3 \Cone \dot Q= -\int \d t \, \d^3 x\, a^3  \frac{H}{c_s^2} \Vone \,.
 \eea
    \noindent  \textbullet \, Performing a temporal integration by part, one finds
 \bea
 \int \d t \, \d^3 x\, a^3  \Cfour \dot Q \ddot Q=-\frac12 \int \d t \, \d^3 x\, a \left(H\dot Q^2 \partial^2 Q+\dot Q^2 \partial^2 \dot Q \right) \,.
 \eea
 Using Eq.~\refeq{simplification1} and $ \int \d t \, \d^3 x\, \dot Q^2 \partial^2 \dot Q=-2  \int \d t \, \d^3 x\, \dot Q (\partial \dot Q)^2$, one then finds
 \bea
 \int \d t \, \d^3 x\, a^3 \left(  \Cfour \dot Q \ddot Q - \Cfive \dot Q \right)=- \int \d t \, \d^3 x\, a^3  \frac{H^2}{c_s^2} \Vone \,.
 \eea
   \noindent  \textbullet \, At this stage, only the last line in Eq.~\refeq{L5cubic} remains to be simplified. All other operators have been solely expressed in terms of $\Vone$ and $\Vtwo$. Remarkably, the same is true here: the four operators $\Cfour^2 \ddot Q$, $\Ceight \ddot Q$, $\Cfour \Cfive$ and $\Cnine$ appear in Eq.~\refeq{L5cubic} in precisely the specific combination that can be related to the operators $\Vone$ and $\Vtwo$\footnote{It can be checked that no other combination of these $4$ operators fulfills this property.}. This relation reads:
 \bea
A \equiv  \int \d t \, \d^3 x\, a^3 \left(  ( \Cfour^2-\Ceight) \ddot Q - 2 \left( \Cfour \Cfive - \Cnine \right)+\frac{H^3}{c_s^4} \left( \Vone+c_s^2 \Vtwo  \right)  \right)=0\,.
\label{def-A}
 \eea
To prove it, let us first perform the temporal integration by part 
\bea
 \int \d t \, \d^3 x\, a^3 \left( \Cfour^2-\Ceight \right) \ddot Q&=&  \int \d t \, \d^3 x \, a^3 H \dot Q  \left( \Cfour^2-\Ceight \right) \nonumber \\
 &+& 2 \int \d t \, \d^3 x\,a^3  \frac{\dot Q}{a} \left(   Q^{,ij} \dot Q_{,ij}-(\partial^2 Q)(\partial^2 \dot Q)  \right)\,.
 \label{A-1}
\eea
Making two successive spatial integrations by part, one can show that
\bea
 \int \d t \, \d^3 x\, \frac{1}{a} \dot Q Q^{,ij}  Q_{,ij}&=& \int \d t \, \d^3 x\,a^3 \left[ \Cone \Cfour - \Csix  +\dot Q \Cfour^2 \right]
 \label{A-2}
\eea
where the combination $\Cone \Cfour - \Csix$ can be expressed in terms of $\Vone$ and $\Vtwo$ using Eqs.~\refeq{C1C4-C6}-\refeq{C2C4} and \refeq{simplification2}. Similarly, 
\bea
 \int \d t \, \d^3 x\, \frac{1}{a} \dot Q Q^{,ij}  \dot Q_{,ij}&=& \int \d t \, \d^3 x\, \frac{1}{a} \left[ (\partial^2 \dot Q)\left(  \dot Q_{,i}  Q^{,i}+\dot Q \partial^2 Q \right)-Q^{,i} \dot Q^{,j} \dot Q_{,ij} \right]\,,
 \label{A-3}
\eea
and a simple spatial integration by part gives
\bea
 \int \d t \, \d^3 x\, \frac{1}{a} \dot Q^{,i} \dot Q^{,j} Q_{,ij}&=&- \int \d t \, \d^3 x\, \frac{1}{a} \left[ Q_{,i} \dot Q^{,i} \partial^2 \dot Q+Q_{,j} \dot Q_{,i} \dot Q^{,ij} \right]\,.
 \label{A-4}
\eea
Now inserting the relations \refeq{A-1}-\refeq{A-4} into the expression Eq.~\refeq{def-A}, and using that\\
\noindent $\int \d t \, \d^3 x\, Q^{,i} \dot Q^{,j} \dot Q_{,ij}=-\frac12  \int \d t \, \d^3 x\, (\partial \dot Q)^2 (\partial^2 Q)$, one finally arrives at $A=0$, as announced.\\

\subsection{Results}

As a result of these manipulations, the total third-order action of interest \refeq{L4cubic}-\refeq{L5cubic} can be solely expressed in terms of the two operators $\Vone$ and $\Vtwo$:
\bea
\left( \L_{4}+\L_{5} \right)_{(\rm cubic)} &=&  \int \d t \, \d^3 x\,a^3 \left[ A_{\Vone} \Vone+ \c^2A_{\Vtwo} \Vtwo \right] \,,
\label{S3}
\eea
with
\bea
 A_{\Vone}&=& \frac{H^2 \dot \phi}{ c_s^4} \left( 6  \left(1-4 c_s^2+2 c_s^4\right) \Ffour+3  \left(1-12 c_s^2+16 c_s^4 \right) X \FfourX -2 c_s^2 \left(4-15 c_s^2 \right) X^2 \FfourXX   \right.
\nn
\\
&& 
\left.  +4 c_s^4 X^3   \FfourXXX  \right) + \frac{H^3 \dot \phi^2}{ 3 c_s^4} \left( - 15 \left(1-8c_s^2+4 c_s^4 \right)\Ffive -6  \left(1-22 c_s^2+25 c_s^4 \right) X \FfiveX  \right.
\nn
\\
&& 
\left.   +24 c_s^2 (1-3 c_s^2) X^2 \FfiveXX -8 c_s^4 X^3  \FfiveXXX  \right)    
\label{AO1}
\eea
and
\bea
  A_{\Vtwo}&=& \frac{H^2 \dot \phi}{c_s^4} \left(  2\left(3-4c_s^2\right) \Ffour+ \left( 3-22 c_s^2 \right) X \FfourX  -6 c_s^2 X^2 \FfourXX \right) \nn \\
  &+&   \frac{H^3 \dot \phi^2}{c_s^4}  \left(-5\left(1-2 c_s^2\right)\Ffive-2 \left(1-9 c_s^2 \right)  X \FfiveX +4 c_s^2 X^2 \FfiveXX  \right)  \,.
  \label{AO2}
\eea

The signals generated by these two operators is well known and has been analyzed in detail in Ref.~\cite{Senatore:2009gt} (see also Ref.~\cite{Chen:2006nt}). The corresponding bispectra are of equilateral type, while an ``orthogonal'' shape emerges as a distinct signature whenever the relative coefficient $A_{\Vtwo}/A_{\Vone}$ between the two operators lies in a specific interval (see Refs.~\cite{RenauxPetel:2011dv,RenauxPetel:2011uk,Renaux-Petel:2013ppa} for the first concrete realization of this mechanism). It is clear from the expressions above that the background quantities give one full freedom to span the full two dimensional space of shapes associated to the operators $\mathcal{O}_{1,2}$. While the overall amplitude of the bispectrum signal is dictated by the magnitude of the coefficients $A_{\Vone}$ and $A_{\Vtwo}$, the single number $A_{\Vtwo}/A_{\Vone}$ thus says it all about its shape, under our two mild assumptions of approximate shift symmetry and small $\epsilon_{mix}$.

\section{Discussion} 
\label{Conclusion}

The quest for the most general scalar-tensor theory free of Ostrogradski instabilities has so far been quite a fruitful and interesting one. It has lead to uncover interesting phenomenology and plenty of different directions are in need for further exploration. The unconventional effects of the generalized Horndeski, or $G^3$, Lagrangian of Ref.~\cite{Gleyzes:2014dya} on the coupling with matter have been mentioned, but there's much more, and a detailed analysis of its screening properties is also worth pursuing \cite{future}. 

Here we have considered its non-Gaussian signatures when it is employed as the inflationary Lagrangian, concentrating on the \textit{Generalized} Horndeski interactions (the case of standard Horndeski interactions has already been treated in Ref.~\cite{RenauxPetel:2011sb} where it was shown that the set of independent cubic operators is the same as in the simpler $k$-inflationary theory). We have determined the corresponding bispectrum, to find that, upon the two light assumptions of approximate shift-symmetry and decoupling of the metric fluctuations, it is entirely captured by the two well-known leading-order $k$-inflationary shapes, thus spanning from the so-called equilateral profile to the orthogonal shape. This result is all the more remarkable as the naive cubic action ---  which we have been able to simplify by using the linear equation of motion --- contains a large number of operators whose intricate expressions involve higher order derivatives.

We conjecture that this result stems from the specific structure imposed on the interactions by the Ostrogradski stability requirements --- which get rid of higher order time derivatives --- augmented by the general covariance and the fact that the dispersion relation is linear, which impose the same scheme for spatial derivatives. This is supported by the realization that more generic models in the effective field theory of inflation of Ref.~\cite{Cheung} require a broader set of operators to fully capture the non-Gaussian signal, this even when metric fluctuations are neglected and a shift symmetry is in place\footnote{For example, it can easily be checked that the use of the linear equation of motion does not suffice to express the operator ${\cal O}_{10}$ in Ref.~\cite{Bartolo:2010bj} (respectively its bispectrum) as a linear combination of the two operators $\Vone$ and $\Vtwo$ of Eq.~\refeq{Standard-operators} (respectively of their bispectra).}.

At a technical level, let us note that we could have alternatively studied the extended Horndeski Lagrangian of Ref.~\cite{Gleyzes:2014dya} by starting from its formulation in the unitary (uniform inflaton) gauge. As the manifestly covariant form Eqs.~\refeq{A2}-\refeq{L55} differ from it by boundary terms, this may have simplified the appearance of the naive cubic action. However, the same work performed here would then have been needed to track down redundancies between interacting operators using the linear equation of motion.

Finally, let us note that our work suggests several interesting venues one could pursue: it is legitimate to wonder whether a simplification analogous to what happens here at the level of the cubic action may arise at higher perturbative orders, as suggested by our results. It would also be interesting to study, in the same spirit as in this work, the class of models recently introduced by Gao in Ref.~\cite{Gao:2014soa}, in which the requirement of a linear dispersion relation is abandoned. We plan to return to these questions and to the exploration of the exciting phenomenology of generalized Horndeski theories in a future work \cite{future}.\\

\subsection*{Acknowledgements} 

We would like to thank  Claudia de Rham, Andrew J. Tolley and the authors of Ref.~\cite{Gleyzes:2014dya} for enlightening discussions, and the latter for pointing out that the expressions corresponding to Eqs.~\refeq{L4}-\refeq{L5} were incomplete in the previous version. SRP would like to thank the Center for Cosmology and Particle Physics at New York University for hospitality near completion of this paper. This work was supported by French state funds managed by the ANR
within the Investissements d'Avenir programme under reference
ANR-11-IDEX-0004-02 and in part by DOE  DE-SC0010600.

\appendix

\section{Consistency check}
\label{other-calculation}

In this appendix, we provide a consistency check of our results by computing and simplifying the cubic action by starting from the form Eqs.~\refeq{L4}-\refeq{L5} of the action, which differs from the form we have been using in the body of the paper by boundary terms. Decomposing $\phi$ into its background plus fluctuating part, one obtains, after a long calculation:
\bea
\frac{\L_{4 ({\rm cubic})} }{\Ffour \dot \phi}&=&  -2( \Cone \Cfour- \Csix) -3 H \Cthree+H  \left(3+ 2 \frac{X \FfourX}{\Ffour}\right)  \Ctwo \Cfour -2 H \Ctwo \ddot Q  \nonumber  \\
&+& 8H \left(1+ \frac{X \FfourX}{\Ffour}  \right) \Cone \dot Q -2H \left(3+7  \frac{X \FfourX}{\Ffour}+2 \frac{X^2 \FfourXX}{\Ffour}  \right) \Cfour \dot Q^2   \nonumber \\
&+& 2 H^2 \left(6+24  \frac{X \FfourX}{\Ffour}+15  \frac{X^2 \FfourXX}{\Ffour}+2  \frac{X^3 \FfourXXX}{\Ffour}  \right) \Vone  \nonumber \\
&-& 2 H^2 \left( 5+11\frac{X \FfourX}{\Ffour}+3 \frac{X^2 \FfourXX}{\Ffour}  \right) \Vtwo   \nonumber \\
&+& \left( 1+\frac{X \FfourX}{\Ffour}  \right) \left( \Cfour^2 \dot Q - \Ceight \dot Q \right)
\eea
and
\bea
\frac{\L_{5 ({\rm cubic})} }{\Ffive \dot \phi^2}&=&  4H( \Cone \Cfour - \Csix) +4 H^2 \Cthree  -2 H^2 \left(2+  \frac{X \FfiveX}{\Ffive} \right) \Ctwo \Cfour +2 H^2 \Ctwo \ddot Q   \nonumber  \\
&-& 4H^2 \left(3+2 \frac{X \FfiveX}{\Ffive}  \right) \Cone \dot Q +2H^2 \left(6+9  \frac{X \FfiveX}{\Ffive}+2 \frac{X^2 \FfiveXX}{\Ffive}  \right) \Cfour \dot Q^2  
\nonumber \\
&-& \frac23 H^3 \left(30+75  \frac{X \FfiveX}{\Ffive}+36  \frac{X^2 \FfiveXX}{\Ffive}+4  \frac{X^3 \FfiveXXX}{\Ffive}  \right) \Vone  \nonumber \\
&+& 2 H^3 \left( 6+9\frac{X \FfiveX}{\Ffive}+2 \frac{X^2 \FfiveXX}{\Ffive}  \right) \Vtwo   \nonumber \\
&-&H \left( 3+2\frac{X \FfiveX}{\Ffive}  \right) \left( \Cfour^2 \dot Q - \Ceight \dot Q \right)  \nonumber \\
&+& \frac13 \left(2 \,\Cten + \Cfour^3 -3\,  \Cfour \Ceight \right)
\,,
\eea
where we defined
\bea
 a^6 \Cten &=& Q_{,ik} Q^{,kj} Q^{,i}_{\,\,j}\,.
\eea
To simplify the above cubic action, one can use the various integrations by part and redundancies used in the subsection \ref{simplify}. Additionally, we need the following relations: \\

  \noindent  \textbullet \, Equation \refeq{A-2} reads
  \bea
 \int \d t \, \d^3 x\, a^3 \dot Q \Ceight &=& \int \d t \, \d^3 x\,a^3 \left[ \Cone \Cfour - \Csix  +\dot Q \Cfour^2 \right]\,,
\eea
where the combination $\Cone \Cfour - \Csix$, and henceforth the combination $ \Cfour^2 \dot Q - \Ceight \dot Q$ as well, can be expressed in terms of $\Vone$ and $\Vtwo$ using Eqs.~\refeq{C1C4-C6}-\refeq{C2C4} and \refeq{simplification2}. \\

  \noindent  \textbullet  \, The combination $2 \,\Cten + \Cfour^3 -3\,  \Cfour \Ceight$ is zero up to boundary terms. To prove this, we integrate by part to write
  \be
   \int  \d^3 x\, 2 \,a^6 \,\Cten= -2 \int  \d^3 x\,\left( Q_{,k} Q^{,kj}_{i} Q^{,i}_{j} + Q_{,k} Q^{,kj} (\partial^2 Q)_{,j} \right)
  \ee
  where
 \be
    \int  \d^3 x\,  Q_{,k} Q^{,kj}_{i} Q^{,i}_{j}=    - \frac12\int  \d^3 x\,  Q_{,ij} Q^{,ij} \partial^2 Q\,.
 \ee 
Writing
\be
    \int  \d^3 x\, (\partial^2 Q)^3=-2     \int  \d^3 x\, Q_{,i} (\partial^2 Q)^{,i} \partial^2 Q\,,
\ee
one thus obtains
\be
  \int  \d^3 x\, a^6 \left(2 \,\Cten + \Cfour^3 -3\,  \Cfour \Ceight \right)= -2 \int  \d^3 x\,\left(Q_{,ij} Q^{,ij} \partial^2 Q+Q_{,i} (\partial^2 Q)^{,i} \partial^2 Q+Q_{,k} Q^{,kj} (\partial^2 Q)_{,j}    \right)\,,
\ee
which a final integration by part of the last term proves to be zero.\\

Using these simplifications and the ones listed in the main body of the paper, one arrives at an expression of the cubic action in terms of the operators $\Vone$ and $\Vtwo$ only that agrees with the result Eqs.~\refeq{S3}-\refeq{AO2}.

\end{document}